\documentclass[11pt,a4paper]{article}

\usepackage[utf8]{inputenc}
\usepackage[T1]{fontenc}
\usepackage[british]{babel}
\usepackage{amsmath, amssymb, amsthm}
\usepackage{mathtools}
\usepackage{xcolor}
\usepackage{listings}
\usepackage{tikz}
\usepackage{geometry}
\usepackage{booktabs}
\usepackage{microtype}
\usepackage{needspace}
\usepackage[hyphens]{url}
\usepackage{xurl}
\usepackage{hyperref}

\hypersetup{
    colorlinks=true,
    linkcolor=black,
    citecolor=blue!50!black,
    urlcolor=blue!50!black
}

\usetikzlibrary{shapes,arrows.meta,positioning,calc,fit}

\lstdefinelanguage{Lean4}{
    morekeywords={theorem, lemma, def, by, cases, intro, refine, simp, simpa,
        apply, exact, have, show, change, unfold, by_cases, classical, with,
        intros, fun, let, match, if, then, else, structure, namespace, end,
        open, import, where, in, do, return, pure, this, Type, Prop, Nat, Int,
        Bool},
    keywordstyle=\color{blue!70!black}\bfseries,
    commentstyle=\color{green!50!black}\itshape,
    stringstyle=\color{red!60!black},
    morecomment=[l]{--},
    morestring=[b]",
    basicstyle=\footnotesize\ttfamily,
    breaklines=true,
    breakatwhitespace=false,
    breakindent=2em,
    postbreak=\mbox{\textcolor{gray}{$\hookrightarrow$}\space},
    columns=fullflexible,
    keepspaces=true,
    showstringspaces=false,
    frame=single,
    numbers=left,
    numberstyle=\tiny\color{gray},
    xleftmargin=1em,
    xrightmargin=1em,
    literate=%
        {α}{{$\alpha$}}1 {β}{{$\beta$}}1 {γ}{{$\gamma$}}1 {Σ}{{$\Sigma$}}1
        {→}{{$\to$}}1 {←}{{$\leftarrow$}}1 {↔}{{$\leftrightarrow$}}1
        {≤}{{$\leq$}}1 {≥}{{$\geq$}}1 {≠}{{$\neq$}}1 {⟨}{{$\langle$}}1
        {⟩}{{$\rangle$}}1 {∃}{{$\exists$}}1 {∀}{{$\forall$}}1 {∧}{{$\wedge$}}1
        {∨}{{$\vee$}}1 {¬}{{$\neg$}}1 {≡}{{$\equiv$}}1 {ℕ}{{$\mathbb{N}$}}1
        {ℤ}{{$\mathbb{Z}$}}1 {ℝ}{{$\mathbb{R}$}}1 {·}{{$\cdot$}}1
        {⁻}{{$^{-}$}}1 {²}{{$^{2}$}}1
        {'}{{\textquotesingle}}1
}

\lstdefinelanguage{RustLang}{
    morekeywords={pub, fn, let, match, if, else, return, struct, enum, impl,
        trait, use, mod, as, mut, ref, in, for, while, loop, break, continue,
        true, false, Some, None, Ok, Err, Option, Result, usize, u8, u16, u32,
        u64, u128, i8, i16, i32, i64, i128, bool, Self, self},
    keywordstyle=\color{orange!80!black}\bfseries,
    commentstyle=\color{green!50!black}\itshape,
    stringstyle=\color{red!60!black},
    morecomment=[l]{//},
    morecomment=[s]{/*}{*/},
    morestring=[b]",
    basicstyle=\footnotesize\ttfamily,
    breaklines=true,
    breakatwhitespace=false,
    breakindent=2em,
    postbreak=\mbox{\textcolor{gray}{$\hookrightarrow$}\space},
    columns=fullflexible,
    keepspaces=true,
    showstringspaces=false,
    frame=single,
    numbers=left,
    numberstyle=\tiny\color{gray},
    xleftmargin=1em,
    xrightmargin=1em
}

\title{A Rust-to-Lean Verification Pipeline with AI Provers:\\
An Experience Report}

\author{
  Natalia Klaus \and
  Juan Conejero \and
  Palina Tolmach \\[0.5em]
  \small Runtime Verification, Inc.\\
  \small \{nat.klaus, juan.conejero, palina.tolmach\}@runtimeverification.com
}

\date{May 2026}

\begin{document}
\maketitle

\begin{abstract}
We describe a verification pipeline that takes production Rust cryptographic
code and produces machine-checked correctness proofs in Lean~4. The pipeline
combines three components: symbolic extraction tools (Charon and Aeneas, or
Hax) that lift Rust into Lean~4; formal cryptographic specification libraries
(ArkLib and CompPoly, from the Verified zkEVM project) that provide the
mathematical targets; and AI provers (Aristotle from Harmonic AI and Aleph
from Logical Intelligence) that close the resulting proof obligations. Every
proof is checked by the Lean kernel, so AI output cannot compromise
soundness.

Within the scope of the Ethereum Foundation's zkEVM Verification Project, we
applied the pipeline to cryptographic primitives in
Plonky3~\cite{plonky3} (FRI folding, Mersenne31 and KoalaBear field
arithmetic, Horner polynomial evaluation) and RISC~Zero (Merkle inclusion
verification). In addition, Aleph authored proofs of two bounds-style
theorems in Plonky3's \texttt{compute\_log\_arity\_for\_round} that
previously stood as \texttt{sorry}.

The paper describes the architecture, walks through a running example based
on Aleph's two proofs, reports which classes of proof obligations AI closed
and which required manual work, and discusses the engineering gaps we
encountered: Lean~4 toolchain drift across tools and specific Aeneas/Hax
extraction limits. We also document concrete missing lemmas, tactic gaps,
and code-generation friction points discovered during proof development. We
hope this contribution lowers the barrier to adoption of formal verification
and facilitates more effective use of AI in this pipeline. The result is a
working pipeline for formal verification of Rust, with kernel-checked
proofs and reproducible artefacts.
\end{abstract}

%% ----------------------------------------------------------------
\section{Introduction}

Cryptographic code that runs inside production protocol stacks tolerates
very few mistakes. A bug in one round of FRI folding silently corrupts
every proof a zero-knowledge virtual machine (zkVM) emits. A bug in a
Merkle inclusion check breaks every fraud proof that depends on it. Code
review, fuzzing, and property-based testing each catch a different slice of
the bug distribution, but each has structural limits. Code review and
property-based testing depend on the auditor or generator anticipating the
right patterns. Fuzzing explores inputs more broadly, but it runs on a
finite input budget and detects only observable failures such as crashes,
hangs, or sanitizer trips. Silent semantic divergences that satisfy these
signals can pass through unnoticed. Formal verification produces a
machine-checked proof covering every input, but only against the property
the engineer has specified. Specification design therefore becomes a
central engineering activity, and one we discuss in Section~\ref{sec:pipeline}.

The cost of formal verification has historically been the obstacle.
Landmark systems such as seL4~\cite{klein2009sel4} and
CompCert~\cite{leroy2009compcert} demonstrated that production-quality
software can be fully verified, but each represents many person-years of
effort in Isabelle/HOL or Rocq, with proofs developed in close coupling to
the implementation. For Rust specifically, recent verifiers such as
Verus~\cite{lattuada2023verus}, Prusti~\cite{astrauskas2022prusti},
Kani~\cite{kani}, and Creusot~\cite{denis2022creusot} have lowered the
entry barrier by embedding specifications directly in Rust source code and
automating common proof obligations through SMT solvers and model
checking. However, they target different fragments of the language and
address different correctness questions, and none of them connects
production Rust to the kind of abstract mathematical specifications that
cryptographic protocols are usually defined against.

Three recent developments together lower the cost of industrial formal
verification substantially. The first is a set of Rust-to-proof-assistant
translation tools that preserve enough of the source semantics to support
real verification: Aeneas~\cite{ho2022aeneas} lowers safe Rust into pure
functional Lean~4 via a typed intermediate representation;
Hax~\cite{hax} (Cryspen) is an annotation-driven alternative targeting
Lean~4, F$^\star$, and Rocq; and rocq-of-rust~\cite{rocqofrust} translates
Rust into Rocq via THIR.

The second is the emergence of AI provers that can close non-trivial proof
obligations in interactive theorem provers:
Aristotle~\cite{achim2025aristotle} from Harmonic AI reports
gold-medal-equivalent performance on the 2025 International Mathematical
Olympiad, and Aleph from Logical Intelligence reports 99.4\% on the
PutnamBench benchmark~\cite{isenbaev2026aleph}. Both produce Lean proofs
that the kernel re-checks, so soundness does not depend on whether the AI
guessed correctly.

The third is the appearance of formal libraries in Lean~4 that make
production verification practical. The foundation is Mathlib4, the
community-maintained mathematical library~\cite{mathlib4}, and
CSLib~\cite{barrett2026cslib}, a recent effort to do for computer science
what Mathlib does for mathematics. Building on these foundations,
specialised libraries provide the abstract targets for individual
verification domains. The two most relevant to this work are
ArkLib~\cite{arklib}, a library for formally verifying SNARK protocols
built on the theory of Interactive Oracle Reductions (with formalisations
of Sum-Check, FRI, WHIR, and others), and CompPoly~\cite{comppoly}, a
library for computational polynomial and finite-field theory. Together,
these libraries provide the abstract mathematical targets that production
Rust can be proved equivalent to.

To the best of our knowledge, this paper documents the first attempt to
combine these three threads in a single pipeline and apply it to
real-world cryptographic Rust. We built and operated such a pipeline under
the Ethereum Foundation's zkEVM Verification Project, applying it to
cryptographic primitives in Plonky3~\cite{plonky3} and
RISC~Zero~\cite{risczero}.

This paper makes four contributions:
\begin{itemize}
  \item \emph{A pipeline.} We integrate symbolic Rust-to-Lean extraction
    (Aeneas, Hax), formal cryptographic specification libraries (ArkLib,
    CompPoly), and AI provers (Aristotle, Aleph) into a single workflow.
  \item \emph{Application to cryptographic primitives.} We apply the
    pipeline to eight consensus-critical targets across
    Plonky3~\cite{plonky3} and RISC~Zero~\cite{risczero}, spanning
    finite-field arithmetic, FRI folding and round-scheduling, Horner
    polynomial evaluation, Merkle inclusion verification, and a 32-bit
    ADC at the bit-vector level. All code, specifications, and proofs are
    public; the main repository is
    \url{https://github.com/Verified-zkEVM/rust-lean}.
  \item \emph{An empirical account of AI-closed proofs.} We report at the
    lemma level which obligations Aleph closed automatically (two bounds
    theorems in \texttt{compute\_log\_arity\_for\_round}) and the proof
    strategy each used.
  \item \emph{Engineering gaps.} We document the gaps we encountered:
    Lean~4 toolchain drift across tools, and specific extraction limits
    in Aeneas and Hax. We describe two reliable workarounds we used in
    practice (fixing the gap upstream and rewriting the function as an
    extraction-friendly model) and briefly discuss layered tooling as a
    direction for unsafe Rust, which our pipeline does not currently
    cover.
\end{itemize}

%% ----------------------------------------------------------------
\section{Background}\label{sec:background}

This section introduces the components we compose into the pipeline.
Readers familiar with Aeneas, Hax, Lean~4, ArkLib, or the AI provers can
skip the corresponding paragraphs.

\subsection{Charon and Aeneas}

Charon~\cite{charon} translates safe Rust into LLBC~\cite{ho2022aeneas}, a
typed intermediate representation derived from Rust's MIR.
Aeneas~\cite{ho2022aeneas} takes LLBC and produces pure functional Lean~4
code. Each Rust function becomes a Lean function in the
\texttt{Result\ $\alpha$} monad, whose three constructors \texttt{ok},
\texttt{fail}, \texttt{div} distinguish successful return, panic, and
non-termination.

Aeneas relieves the proof engineer of memory-based
reasoning~\cite{ho2022aeneas}: it treats Rust's borrow-checker guarantees
as a semantic input. The extracted Lean code stays small and focused on
functional properties, provable with Mathlib4 and Aeneas-specific tactics
such as \texttt{progress} and \texttt{scalar\_tac}.

\subsection{Hax}

Hax~\cite{hax} is a Rust verification tool maintained by Cryspen that
translates Rust into F$^\star$, Rocq, Lean, and other backends. Its
frontend hooks into the Rust compiler and exports the THIR (Typed
High-level Intermediate Representation) as JSON; the engine then lowers
this through an annotation-driven simplification pipeline. The Lean~4
backend is currently under active development, while the F$^\star$ backend
is more mature.

Recent independent analysis~\cite{symbolic2026hax} has identified semantic
gaps between Hax extractions and the original Rust semantics for certain
Rust patterns; the verification work reported here stays within the
fragment of pure, bounded-loop functions where Hax's translation is
reliable.

\subsection{Lean 4 and the spec libraries}\label{sec:lean-spec-libs}

The target prover is Lean~4 with Mathlib4 as its mathematical library. Two
further libraries provide the cryptographic specifications we verify
against:

ArkLib~\cite{arklib} formalises succinct non-interactive arguments of
knowledge (SNARKs), cryptographic proof systems that allow a prover to
convince a verifier that a computation was performed correctly without the
verifier re-executing it. ArkLib builds on the theory of Interactive
Oracle Reductions (IORs), a compositional framework in which a complex
proof system is decomposed into a sequence of simpler interactive
protocols, each reducing one relation to another~\cite{chiesa2019ior}. The
library includes formalisations of Sum-Check, FRI, WHIR, and other
protocols at the abstract level, together with the polynomial machinery
(folding, evaluation domains) they depend on.

CompPoly~\cite{comppoly} provides computational polynomial and
finite-field theory, including univariate, multivariate, and multilinear
polynomial representations with equivalences to Mathlib's algebraic
types, used both directly and as a foundation under ArkLib.

Both libraries are open-source and part of the Verified zkEVM project. As
part of this work, we have contributed upstream improvements to both
libraries.

\subsection{AI provers}

We use two AI provers as external proof-search engines for Lean~4.
Aristotle (Harmonic AI)~\cite{achim2025aristotle} combines a Lean
proof-search system with an informal reasoning component that generates
and formalises candidate lemmas. Aleph (Logical
Intelligence)~\cite{aleph-site} is an agentic orchestration layer for
theorem proving: it decomposes a proof obligation into subproblems
(``planning''), generates Lean proofs for each (``proving''), and refines
its strategy based on which subparts succeed (``refining''). Aleph can be
paired with different underlying reasoning models depending on the task
and resource budget. The Lean kernel re-checks every proof both provers
emit.

\subsection{zkVMs and Plonky3}

Zero-knowledge virtual machines (zkVMs) execute programs inside a
cryptographic proof system: the VM produces a succinct proof that the
execution was correct, which any third party can verify without
re-executing the program. RISC~Zero~\cite{risczero} and
SP1~\cite{sp1} are two production zkVMs built on the RISC-V instruction
set. Both rely on STARK proof systems~\cite{bensasson2018starks} whose
arithmetic backbone is implemented in Plonky3~\cite{plonky3}.

Plonky3~\cite{plonky3} is an open-source Rust toolkit for building STARK
proof systems. It provides finite field implementations (Mersenne31,
BabyBear, KoalaBear), the Fast Reed-Solomon Interactive Oracle Proof of
Proximity (FRI) protocol~\cite{bensasson2018starks}, and the polynomial
commitment infrastructure that zkVMs build on. FRI is a low-degree test
that lets a verifier check whether a function evaluated over a finite
field is close to a low-degree polynomial. It does so through a
logarithmic number of folding rounds, each reducing the problem size by
half. A bug in any of these components, such as a wrong field
multiplication, an incorrect FRI folding step, or a flawed Merkle
inclusion check, silently invalidates every proof the zkVM emits.

The Verified zkEVM project~\cite{zkevm-project} is a multi-team effort to
formally verify components across this stack. ArkLib and CompPoly
(Section~\ref{sec:lean-spec-libs}) were developed as part of this project. The
verification work reported in this paper was carried out under the same
project, targeting Plonky3 and RISC~Zero primitives as verification
subjects.

%% ----------------------------------------------------------------
\section{Pipeline Architecture}\label{sec:pipeline}

\begin{figure}[t]
\centering
\begin{tikzpicture}[
    node distance=7mm and 12mm,
    every node/.style={font=\small,align=center},
    box/.style={rectangle,rounded corners=2mm,draw,thick,minimum width=42mm,minimum height=8mm,inner sep=2pt},
    extract/.style={box,fill=black!8},
    spec/.style={box,fill=green!10},
    proof/.style={box,fill=blue!10},
    kernel/.style={box,fill=red!12,line width=0.7mm},
    aux/.style={rectangle,rounded corners=2mm,draw,minimum width=38mm,minimum height=7mm,fill=white,inner sep=2pt,font=\footnotesize},
    arr/.style={-Stealth,thick}
  ]
  \node[extract] (rust) {Rust crate\\[-1pt]\scriptsize (Plonky3, RISC Zero, \ldots)};
  \node[extract,below=of rust] (charon) {Charon\\[-1pt]\scriptsize (Rust $\to$ LLBC)};
  \node[extract,below=of charon] (extract) {Aeneas / Hax\\[-1pt]\scriptsize (LLBC $\to$ Lean 4)};
  \node[extract,below=of extract] (lean) {Lean 4 extracted code};
  \node[proof,below=of lean] (thm) {Theorem statement};
  \node[proof,below=of thm] (pf) {Proof construction};
  \node[kernel,below=of pf] (k) {Lean 4 kernel\\[-1pt]\scriptsize re-checks every proof};
  \node[extract,below=of k] (out) {Machine-checked theorems};

  % Specifications side
  \node[spec,right=22mm of thm] (specs) {Specifications};
  \node[aux,below right=2mm and -3mm of specs,fill=green!5] (hand) {Handwritten specs};
  \node[aux,below=1mm of hand,fill=green!5] (arklib) {ArkLib / CompPoly};

  % Proof side (placed below arklib so the spec arrows don't cross through them)
  \node[aux,below=6mm of arklib,fill=blue!5,minimum width=42mm] (manual) {Manual proof};
  \node[aux,below=1mm of manual,fill=blue!5,minimum width=42mm] (ai) {AI provers\\[-1pt]\scriptsize (Aristotle, Aleph)};

  \draw[arr] (rust) -- (charon);
  \draw[arr] (charon) -- (extract);
  \draw[arr] (extract) -- (lean);
  \draw[arr] (lean) -- (thm);
  \draw[arr] (thm) -- (pf);
  \draw[arr] (pf) -- (k);
  \draw[arr] (k) -- (out);

  \draw[arr] (specs.west) -- (thm.east);
  \draw[arr] (hand.west) to[bend left=8] ($(thm.east) + (0,-1mm)$);
  \draw[arr] (arklib.west) to[bend left=12] ($(thm.east) + (0,-2mm)$);

  \draw[arr] (manual.west) -- (pf.east);
  \draw[arr] (ai.west) to[bend left=10] ($(pf.east) + (0,-2mm)$);
\end{tikzpicture}
\caption{Pipeline data flow. Extraction (grey) lifts Rust into Lean 4;
specifications (green) define the verification targets; proofs (blue) are
constructed manually and with AI assistance; the Lean kernel (red) re-checks
every proof, constituting the trust boundary.}
\label{fig:pipeline}
\end{figure}
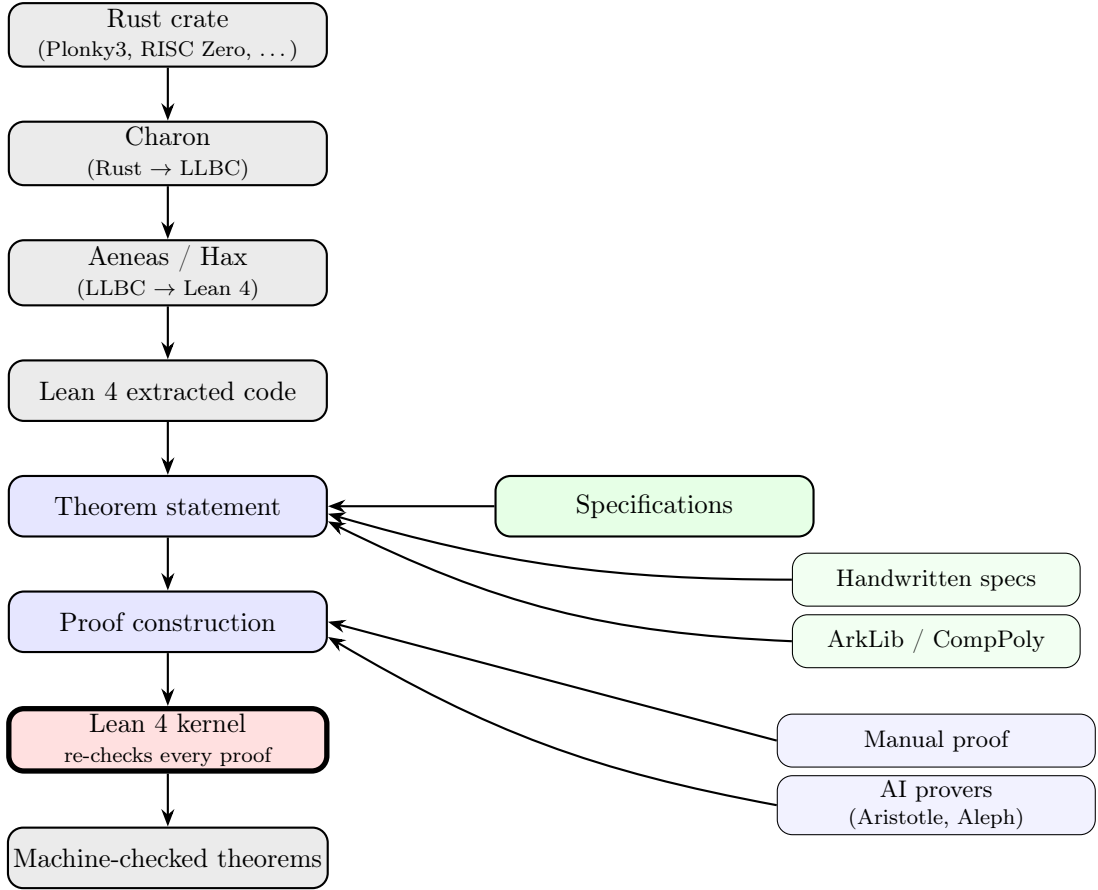

Figure~\ref{fig:pipeline} shows the data flow. The pipeline runs in four
stages.

\paragraph{Stage 1: Rust to Lean 4.}
For projects using Aeneas, we run Charon on the Cargo crate to produce an
LLBC file, then run Aeneas to translate LLBC into a pair of Lean files
(\texttt{Types.lean} and \texttt{Funs.lean}). For projects using Hax, we
annotate Rust items with \texttt{\#[hax::contract]} and related attributes,
then run \texttt{cargo hax into lean}.

In both cases, the output is pure functional Lean~4. Rust loops become
recursive functions with termination obligations. Arithmetic operations
are wrapped in monads (\texttt{Result\ $\alpha$} for Aeneas,
\texttt{RustM} for Hax) that expose overflow and panic paths as explicit
failure cases.

\paragraph{Stage 2: Specification.}
Specifications come from two sources. For simple targets, we write them by
hand against the extracted code: a pure function (over $\mathbb{N}$ or
$\mathbb{Z}/p\mathbb{Z}$) that mirrors the Rust computation, together with
preconditions on the inputs (for example, $p^2 \leq 2^{64}-1$, or
``all inputs are less than $p$''). For targets that match an existing
cryptographic abstraction, we import the relevant definition from ArkLib
or CompPoly and state the spec as equivalence to that definition. The
FRI fold example in Section~\ref{sec:running} takes this second route:
the Rust output is proved equal to one evaluation of ArkLib's
\texttt{foldNth 2 f $\beta$} polynomial fold.

The main theorem of each verification target then states: under the
precondition structure, the extracted Rust function returns a value
matching the specification.

\paragraph{Stage 3: Proof.}
Proofs are constructed either manually or with the help of AI provers, and
typically combine both. The available automation includes:
\begin{itemize}
  \item Aeneas-specific tactics (when extraction goes through Aeneas):
    \texttt{progress} steps through monadic \texttt{let\ x\ $\leftarrow$\ e}
    bindings and closes ``operation succeeds'' obligations using
    preconditions in context; \texttt{scalar\_tac} discharges integer-bound
    side conditions.
  \item General Lean~4 tactics: \texttt{omega} (linear arithmetic),
    \texttt{nlinarith} (nonlinear bounds), \texttt{ring} (algebraic
    identities), \texttt{simp} (rewriting), and \texttt{decide} (decidable
    propositions). These come from Lean~4 core and Mathlib4.
  \item AI provers: Aristotle and Aleph can be invoked to close individual
    lemmas or whole theorems. Section~\ref{sec:ai} gives an empirical
    account of which classes of obligation they handle well.
\end{itemize}
Manual and AI-driven proofs coexist in the same Lean file. There is no
semantic distinction between them at the kernel level.

\paragraph{Stage 4: Re-check.}
The Lean kernel re-checks every proof, constituting the trust boundary of
the described pipeline: regardless of whether a proof was written by a
human, generated by an AI prover, or assembled from automated tactics,
the kernel accepts it only if it typechecks.

%% ----------------------------------------------------------------
\section{Running Example: \texttt{compute\_log\_arity\_for\_round}}\label{sec:running}

We use Plonky3's \texttt{compute\_log\_arity\_for\_round}, an
FRI~\cite{bensasson2018starks} round-scheduling function from
\texttt{fri/src/config.rs}, as the running example. The function decides
how aggressively a given FRI round folds, subject to three constraints: a
maximum allowed arity, the distance to the final target height, and the
distance to the next input commitment.

The Plonky3 source, lightly elided:

\needspace{20\baselineskip}
\begin{lstlisting}[language=RustLang]
pub fn compute_log_arity_for_round(
    log_current_height: usize,
    next_input_log_height: Option<usize>,
    log_final_height: usize,
    max_log_arity: usize,
) -> RustM<usize> {
    let max_fold_to_target = log_current_height - log_final_height;
    let max_fold = match next_input_log_height {
        None => max_fold_to_target,
        Some(next) => {
            let to_next = log_current_height - next;
            if to_next < max_fold_to_target { to_next } else { max_fold_to_target }
        }
    };
    if max_fold < max_log_arity { max_fold } else { max_log_arity }
}
\end{lstlisting}

Hax extracts this into Lean~4, wrapping each checked subtraction in a
\texttt{RustM} monad and representing each conditional as a Lean
\texttt{if/then/else}. We can formulate two theorems: the result must not
exceed the maximum allowed arity (\texttt{arity\_respects\_max\_bound}),
and the result must not exceed the distance to the final height
(\texttt{arity\_respects\_target\_distance}). Both bounds are necessary
for the soundness of every subsequent FRI round. Both initially stood as
\texttt{sorry} in
\texttt{lean/P3FriProofs/Proofs/FoldingCorrectness.lean}. Both were closed
in February 2026 by the \texttt{aleph-prover[bot]} GitHub account,
running Aleph against the Lean files.

\subsection{Aleph's proof of \texttt{arity\_respects\_max\_bound}}\label{sec:proof-max-bound}

PR \#1 to \texttt{p3-hax-lean-fri-pipeline}~\cite{alephpr1} was merged on
23 February 2026. The PR description, generated by Aleph, identifies the
following proof strategy: decompose into two reusable helper lemmas about
the Hax-extracted \texttt{RustM} monad, peel the monadic binds of
\texttt{compute\_log\_arity\_for\_round} one by one, and reduce to the
final ``min with cap'' \texttt{if}.

The first helper inverts a successful monadic bind: if
\texttt{x\ >>=\ f\ =\ ok\ r}, then there exists some \texttt{a} with
\texttt{x\ =\ ok\ a} and \texttt{f\ a\ =\ ok\ r}.

\needspace{12\baselineskip}
\begin{lstlisting}[language=Lean4]
theorem RustM_bind_eq_ok {α β : Type} (x : RustM α) (f : α → RustM β) (r : β) :
    RustM.bind x f = .ok r → ∃ a, x = .ok a ∧ f a = .ok r := by
  cases x with
  | ok a =>
    intro hx; refine ⟨a, rfl, ?_⟩
    simpa [RustM.bind] using hx
  | fail e => intro hx; simp [RustM.bind] at hx
  | div => intro hx; simp [RustM.bind] at hx
\end{lstlisting}

The second helper derives a bound from the ``min'' if-expression: if
\texttt{(if a < b then ok a else ok b) = ok r}, then \texttt{r $\leq$ b}.

\needspace{14\baselineskip}
\begin{lstlisting}[language=Lean4]
theorem RustM_ok_ite_decide_lt_le_right (a b r : USize64) :
    (if decide (a < b) then (RustM.ok a : RustM USize64) else RustM.ok b) =
      RustM.ok r → r ≤ b := by
  intro h
  by_cases hlt : a < b
  · simp [hlt] at h; cases h
    have hnat : a.toNat < b.toNat := by simpa using hlt
    change a.toNat ≤ b.toNat
    exact Nat.le_of_lt hnat
  · simp [hlt] at h; cases h; change b.toNat ≤ b.toNat; exact le_rfl
\end{lstlisting}

The main theorem unfolds \texttt{compute\_log\_arity\_for\_round} and uses
\texttt{RustM\_bind\_eq\_ok} to peel off the first checked subtraction. It
then case-splits on \texttt{next\_input\_log\_height}, peels the second
bind in the \texttt{Some} branch, and case-splits on the inner
\texttt{<} comparison. Finally, it applies
\texttt{RustM\_ok\_ite\_decide\_lt\_le\_right} to the
``min-with-cap'' \texttt{if}, bounding the result by
\texttt{max\_log\_arity}. Aleph observes that the arithmetic precondition
\texttt{h\_gt} is unnecessary for this bound; this observation avoids any
checked-subtraction reasoning.

\subsection{Aleph's proof of \texttt{arity\_respects\_target\_distance}}

PR \#3~\cite{alephpr3} was merged the following day, on 24 February 2026,
with a symmetric strategy. A single helper derives that
\texttt{(if a < b then ok a else ok b) = ok r} implies \texttt{r $\leq$ a}:

\needspace{13\baselineskip}
\begin{lstlisting}[language=Lean4]
theorem RustM_ok_ite_decide_lt_le_left (a b r : USize64) :
    (if decide (a < b) then (RustM.ok a : RustM USize64) else RustM.ok b) =
      RustM.ok r → r ≤ a := by
  intro h
  by_cases hlt : a < b
  · simp [hlt] at h; cases h; change a.toNat ≤ a.toNat; exact le_rfl
  · simp [hlt] at h; cases h
    change b.toNat ≤ a.toNat
    apply Nat.le_of_not_gt; simpa using hlt
\end{lstlisting}

The main theorem is more involved than in
Section~\ref{sec:proof-max-bound}. The conclusion
\[
  \mathtt{result.toNat}
    \leq
  \mathtt{log\_current\_height.toNat}
    -
  \mathtt{log\_final\_height.toNat}
\]
involves checked subtraction on \texttt{USize64}: at the Rust level the
subtraction would have underflowed if \texttt{log\_final\_height}
$\geq$ \texttt{log\_current\_height}, and the proof must discharge that
case using the height-inequality hypothesis \texttt{h\_gt}.

\needspace{28\baselineskip}
\begin{lstlisting}[language=Lean4]
theorem arity_respects_target_distance
    (log_current_height log_final_height max_log_arity result : USize64)
    (h_gt : log_current_height.toNat > log_final_height.toNat)
    (h_result : P3_fri_kernel.compute_log_arity_for_round
        log_current_height Core_models.Option.Option.None
        log_final_height max_log_arity = .ok result)
    : result.toNat ≤ log_current_height.toNat - log_final_height.toNat := by
  classical
  have hgt' : log_final_height.toNat < log_current_height.toNat := by simpa using h_gt
  have hnot : ¬ (log_current_height.toNat < log_final_height.toNat) :=
    Nat.not_lt_of_ge (Nat.le_of_lt hgt')
  unfold P3_fri_kernel.compute_log_arity_for_round at h_result
  simp [RustM.bind, Core_models.Ops.Arith.Sub.sub, instSubUSize64_1,
        BitVec.usubOverflow, hnot] at h_result
  have hle : result ≤ log_current_height - log_final_height := by
    apply RustM_ok_ite_decide_lt_le_left
      (a := log_current_height - log_final_height) (b := max_log_arity) (r := result)
    by_cases hlt : log_current_height - log_final_height < max_log_arity <;>
      simpa [hlt] using h_result
  have hleNat : result.toNat ≤ (log_current_height - log_final_height).toNat :=
    (USize64.le_iff_toNat_le).1 hle
  have hsub : (log_current_height - log_final_height).toNat =
              log_current_height.toNat - log_final_height.toNat :=
    USize64.toNat_sub_of_le' (Nat.le_of_lt hgt')
  simpa [hsub] using hleNat
\end{lstlisting}

Both PRs are notable for how Aleph closed the theorems: by identifying and
proving small reusable helper lemmas, then composing them in a
structurally clean main proof. The proof-plan blueprints in the PR
descriptions read like a proof engineer's notes; for example,
``decomposition into a key helper lemma and the main bound\ldots\ the
proof must use \texttt{h\_gt} to justify that the subtraction behaves like
ordinary $\mathbb{N}$ subtraction on \texttt{toNat}\ldots\ the min/if
structure is handled cleanly by a \texttt{by\_cases} split and
\texttt{simp}.''

%% ----------------------------------------------------------------
\section{AI-Driven Proof Automation: Empirical Notes}\label{sec:ai}

This section reports what we observed when running Aristotle and Aleph
against the proof obligations that arose in our verification work. We
recognise the small size of our data set: two theorems that Aleph closed
(PR \#1 and PR \#3), plus informal experiments on neighbouring lemmas
during development. Even so, we observed a consistent pattern throughout
all our experiments.

\paragraph{What AI provers handled well.}
Three categories of obligation closed reliably:
\begin{itemize}
  \item \emph{Control-flow and structural lemmas:} monad inversions,
    if-conditional bounds, case-splits on enums. The two PRs of
    Section~\ref{sec:running} fall here. The reusable helper lemmas
    Aleph produced (\texttt{RustM\_bind\_eq\_ok},
    \texttt{RustM\_ok\_ite\_decide\_lt\_le\_left/right}) are typical of
    this category.
  \item \emph{Linear arithmetic with mild side conditions,} particularly
    where \texttt{omega} or \texttt{nlinarith} closes the goal once the
    right hypothesis is named.
  \item \emph{\texttt{simp}-closeable boilerplate} around extracted code:
    unfolding monadic bind, distributing over conditionals, normalising
    checked-arithmetic guards.
\end{itemize}

\paragraph{What required manual work.}
Three categories required either substantial manual proof or full
hand-construction:
\begin{itemize}
  \item \emph{Domain-specific algebraic identities} that require
    carefully selecting Mathlib4 lemmas. A clear example is the
    equivalence between the Aeneas-extracted \texttt{fold\_step} and
    ArkLib's \texttt{foldNth 2 f $\beta$} on $\mathbb{Z}/p\mathbb{Z}$
    polynomials. The obligation is true, but no off-the-shelf Mathlib
    lemma closes it without human direction.
  \item \emph{Loop invariants on Aeneas-extracted recursive functions.}
    We did not observe AI provers bypassing the inherent difficulty of
    finding the right invariant, so this remained a manual task in our
    work.
  \item \emph{Obligations that require axiomatising an external
    interface,} where the choice of axioms is itself a design decision
    rather than a proof-search problem.
\end{itemize}

\paragraph{Implication for the workflow.}
In the work reported here, AI provers operated as a productivity
multiplier on the obligations in the first category, but did not change
the fundamental shape of the verification effort. Specification design,
invariant discovery, and the selection of mathematical abstractions
remained the work of human proof engineers. We expect this division to
shift as AI provers mature, and Section~\ref{sec:conclusion} outlines two
directions where we plan to push that line.

%% ----------------------------------------------------------------
\section{Case Studies}\label{sec:cases}

This section lists the cryptographic targets we verified under the zkEVM
Verification Project. All artefacts are public:
\begin{itemize}
  \item Main verification repository (Aeneas-based work):
    \url{https://github.com/Verified-zkEVM/rust-lean}
  \item Plonky3 FRI pipeline (Hax-based, with AI-prover contributions):
    \url{https://github.com/runtimeverification/p3-hax-lean-fri-pipeline}
  \item Plonky3 FRI fold arity (Aeneas-based):
    \url{https://github.com/runtimeverification/aeneas_fri_fold_arity_verification}
  \item Plonky3 field arithmetic (Aeneas-based):
    \url{https://github.com/runtimeverification/aeneas-field-arithmetic}
  \item 32-bit ADC (Hax-based):
    \url{https://github.com/runtimeverification/hax-annotations-adc-verification}
  \item RISC~Zero Merkle inclusion:
    \url{https://github.com/runtimeverification/zkvm-merkle-lean}
  \item Plonky3 Horner evaluation (Hax + CompPoly):
    \url{https://github.com/runtimeverification/horner_hax_comppoly_poc}
\end{itemize}

\begin{table}[t]
\centering
\small
\begin{tabular}{p{4.4cm} p{2.2cm} p{7.0cm}}
\toprule
\textbf{Target} & \textbf{Pipeline} & \textbf{What was proved} \\
\midrule
Plonky3 FRI round-scheduling
(\texttt{compute\_log\_\allowbreak{}arity\_\allowbreak{}for\_\allowbreak{}round}) &
Aeneas + Hax &
6 theorems incl.\ bounds, monotonicity, boundary cases; 2 closed by Aleph
(PR \#1, PR \#3) \\
Plonky3 arity-2 FRI fold (\texttt{fold\_step}) &
Aeneas + ArkLib &
Rust implementation matches abstract \texttt{foldNth 2 f $\beta$}; no
overflow across 13 checked operations \\
Plonky3 Mersenne31 field arithmetic & Aeneas &
Addition and multiplication of the Rust model \\
Plonky3 Mersenne31 / KoalaBear fields & Hax &
Extraction of the Rust field implementations \\
Plonky3 Horner polynomial evaluation & Hax + CompPoly &
Evaluation correctness against the polynomial spec \\
Plonky3 one-step FRI folding & Hax &
End-to-end FRI round correctness \\
RISC~Zero Merkle inclusion & Hax &
Root recomputation and inclusion-proof verification \\
32-bit ADC (addition with carry) & Hax + \texttt{bv\_decide} &
Bit-level correctness via automated bit-vector reasoning \\
\bottomrule
\end{tabular}
\caption{Verification targets under the zkEVM Verification Project.}
\label{tab:cases}
\end{table}

\paragraph{Case study 1: \texttt{fold\_step} $\leftrightarrow$ ArkLib.}
The arity-2 FRI folding step takes evaluations $lo = f(x)$ and
$hi = f(-x)$ of a polynomial $f$ over a prime field $F_p$, together with
a challenge $\beta$ and precomputed inverses $x^{-1}, 2^{-1}$, and
computes
$\mathit{fold\_step} = (lo + hi + \beta \cdot (lo - hi) \cdot x^{-1}) \cdot 2^{-1} \bmod p$.
This equals $(\mathsf{foldNth}\,2\,f\,\beta).\mathsf{eval}(x^2)$: one
evaluation of the folded polynomial in ArkLib's definition. The Aeneas
extraction unfolds the Rust into a sequence of 13 checked arithmetic
operations on \texttt{u64}, each returning \texttt{Result}. We proved, in
Lean, that under explicit preconditions on $p$
($p^2 \leq 2^{64}-1$, all inputs strictly less than $p$) no operation
overflows, the final result is in $[0, p)$, and the algebraic expression
matches the closed-form above in $\mathbb{Z}/p\mathbb{Z}$ (verified by
the \texttt{ring} tactic). This is the project's tightest connection from
Aeneas-extracted production Rust to a formal cryptographic specification:
the verifier checks not only that the arithmetic is overflow-free, but
that the Rust implementation computes exactly the function ArkLib
defines.

\paragraph{Case study 2: the running example revisited.}
This case study revisits the function
\texttt{compute\_\allowbreak{}log\_\allowbreak{}arity\_\allowbreak{}for\_\allowbreak{}round}
and its correspondence to Aleph's proofs. It is the running example of
Section~\ref{sec:running}. The Hax
extraction produces a Lean~4 function over the Hax \texttt{RustM} monad
with two checked subtractions, an \texttt{Option} match, and two
if-conditional ``min'' selections. Six bounds-style and boundary-case
theorems were stated. Two were closed automatically by Aleph in PRs \#1
and \#3, each producing a small reusable helper lemma in addition to the
main theorem. The remaining four
(\texttt{arity\_respects\_next\_input},
\texttt{folding\_respects\_final\_height},
\texttt{round\_consistency\_preserved}, and one more) currently remain at
\texttt{sorry} in the public repository, awaiting either a follow-up
Aleph run or manual completion.

\paragraph{Repositories and reproducibility.}
All code, specifications, and proofs in this paper are public:
\begin{itemize}
  \item Aeneas-based verification of Plonky3 FRI primitives:
    \url{https://github.com/Verified-zkEVM/rust-lean}
  \item Hax-based verification with AI-prover contributions:
    \url{https://github.com/runtimeverification/p3-hax-lean-fri-pipeline}
  \item Aleph PRs \#1~\cite{alephpr1} and \#3~\cite{alephpr3} (running
    example).
\end{itemize}
A complete list of per-target repositories appears in the bullet list at
the start of this section.

%% ----------------------------------------------------------------
\section{Engineering Challenges and Solutions}\label{sec:engineering}

\paragraph{Lean 4 version drift.}
Aeneas, Hax, ArkLib, CompPoly, and Mathlib4 evolve on independent Lean~4
release tracks. Early in the project we could not build a single Lake
project that imported tools from more than one source, because each was
pinned to a different Lean toolchain. We raised the issue in the affected
repositories; after coordination across the maintainer teams, the
libraries were aligned on a common Lean~4 version (4.26.0 for the
Hax-based work; 4.28.0 for the Aeneas-based work).

\paragraph{Aeneas/Hax extraction gaps.}
The supported subset of safe Rust that Aeneas and Hax can extract to pure
Lean is finite. We have identified the following gaps in practice:
\begin{itemize}
  \item A name collision (\texttt{BitVec.toNat\_pow}) between Aeneas's
    bit-vector lemmas and Batteries / Mathlib4 prevented importing both
    libraries in a single Lean file. We patched this upstream in
    \texttt{AeneasVerif/aeneas\#832}.
  \item Casts to wider integer types (\texttt{u128}) inside modular
    arithmetic require careful proof engineering: \texttt{progress}
    introduces a side condition that \texttt{omega} cannot close on its
    own and that needs \texttt{nlinarith}.
  \item Generic functions with trait bounds (e.g.,
    \texttt{F: Field + TwoAdicField}) cannot be extracted directly:
    Aeneas needs concrete monomorphised types. We worked around this by
    writing a monomorphic model of the function in question (Plonky3's
    \texttt{fold\_matrix} rewritten as a \texttt{u64}-only
    \texttt{fold\_step}); the model is semantically equivalent to the
    production code and extracts cleanly.
  \item External-crate calls (e.g., \texttt{byteorder},
    \texttt{std::io::Read}/\texttt{Write}) are not extractable because
    Charon does not see the MIR of crates beyond the workspace.
  \item Rust \texttt{while} loops become recursive functions in Lean and
    require explicit termination measures.
\end{itemize}

\paragraph{Three reliable workarounds.}
We converged on three strategies, applied in this order of preference:
\begin{enumerate}
  \item \emph{Fix it upstream.} When a gap corresponds to a missing
    feature or lemma that we are able to contribute, we submit the fix
    to the upstream repository. The \texttt{BitVec.toNat\_pow} fix and
    several CompPoly contributions originated this way.
  \item \emph{Extract a model.} When the production Rust code uses
    patterns Aeneas cannot handle (parallel iterators, deep generics,
    external-crate calls), we rewrite the mathematical core as a
    standalone, non-generic Rust crate. The model is semantically
    equivalent, extracts cleanly, and can be verified end-to-end. This
    is the strategy that produced \texttt{fold\_step} from Plonky3's
    \texttt{fold\_matrix}.
  \item \emph{Layered tools for \texttt{unsafe} code.} The pipeline as
    described handles safe Rust. For codebases with \texttt{unsafe}
    blocks, the verification story is layered: prove the safe core with
    Aeneas or Hax, and bring in complementary tools
    (\texttt{cargo-anneal}~\cite{cargoanneal}) for the
    \texttt{unsafe} parts.
\end{enumerate}

%% ----------------------------------------------------------------
\section{Related Work}\label{sec:related}

\paragraph{Rust formal verification.}
Aeneas and Hax are the two extraction-based approaches we use directly.
The other major points in the verification space are
Verus~\cite{lattuada2023verus} (ghost types and an SMT backend, tightly
integrated with the Rust toolchain), Prusti~\cite{astrauskas2022prusti}
(specification annotations and the Viper verifier), Kani~\cite{kani}
(bounded model checking, complementary to formal proof), and
Creusot~\cite{denis2022creusot} (Why3-based deductive verification of
annotated Rust). These tools target different fragments of Rust and rely
on different proof backends (SMT solvers, Viper, Why3, model checking).
Our work instead focuses on connecting production Rust to Mathlib4-level
formal cryptographic specifications, and on integrating AI provers to
close the resulting proof obligations.

\paragraph{Foundational verified systems.}
seL4 and CompCert are landmark projects that demonstrate that complete
formal verification of production-quality software is achievable, but
each represents many person-years of effort in tightly coupled
implementation and proof pairs in Isabelle/HOL and Rocq. The pipeline
described here requires substantially smaller per-engagement effort, in
exchange for verifying specific critical-core components rather than
whole systems.

\paragraph{AI provers for Lean.}
LeanDojo~\cite{yang2023leandojo} provides infrastructure for training and
evaluating LLM-based provers in Lean. Lean Copilot~\cite{song2024leancopilot}
explores LLMs as interactive proof assistants inside the editor. The
AlphaProof and AlphaGeometry~\cite{trinh2024alphageometry} effort from
DeepMind demonstrated silver-medal-equivalent performance on IMO 2024
using reinforcement-learning-trained Lean provers. Our work uses
Aristotle~\cite{achim2025aristotle} and Aleph~\cite{isenbaev2026aleph}
not as benchmarks on synthetic problems but as black-box proof-search
engines on the proof obligations our extraction pipeline emits. This is,
to our knowledge, among the first publicly visible neurosymbolic
deployments on production zkVM cryptographic code.

\paragraph{Cryptographic specification libraries.}
ArkLib and CompPoly are the spec libraries this work targets. Both are
public and we contribute upstream to them. The closest comparable project
is HACL$^\star$ for verified cryptographic implementations in
F$^\star$, but its target language and specification style differ.

%% ----------------------------------------------------------------
\section{Conclusion and Future Work}\label{sec:conclusion}

We described an end-to-end neurosymbolic pipeline from production Rust to
Lean~4 proofs, applied it to cryptographic primitives in Plonky3 and
RISC~Zero under the Ethereum Foundation zkEVM Verification Project, and
gave an empirical account of where AI provers were and were not
effective. The two-PR running example (Aleph proving
\texttt{arity\_respects\_max\_bound} and
\texttt{arity\_respects\_target\_distance} in
\texttt{p3-hax-lean-fri-pipeline}) shows that current AI provers can
close non-trivial structural lemmas about Hax-extracted Rust monadic
code, with proof plans that are themselves auditable.

Two directions of follow-up are immediate. First, we plan to participate
in the Beneficial AI Foundation's Signal Shot programme~\cite{signalshot},
which targets end-to-end machine-checked verification of the Signal
messaging stack in Lean~4. The Signal codebase is a natural next target
for the same pipeline. Second, we plan to integrate AI provers more
tightly into the verification CI/CD loop. Instead of invoking them per
pull request, we would run them automatically on every freshly introduced
\texttt{sorry} and let human reviewers see which obligations are now
closed before they look at the code.

\paragraph{Acknowledgements.}
This work was funded by the Ethereum Foundation under the zkEVM
Verification Project. We are grateful to Alexander Hicks and Derek
Sorensen for initiating and guiding this engagement, providing feedback
on specifications and proofs, and facilitating the collaboration across
teams. We also thank the maintainers of Aeneas, Charon, Hax, Mathlib4,
ArkLib, CompPoly, Aristotle, and Aleph for their open-source work and for
cross-repository coordination on Lean toolchain alignment.

%% ----------------------------------------------------------------
\bibliographystyle{plain}
\bibliography{references}

\end{document}